\documentclass[useAMS, usenatbib, ]{mn2e}
\usepackage{graphicx} 
\usepackage{epstopdf} 
\usepackage{comment}
\usepackage{tabularx}
\usepackage{fixltx2e} 
\title[DECam View of Carina]{Sailing under the Magellanic Clouds: A DECam View of the Carina Dwarf} 
\author[B. McMonigal et al.]
	{B. McMonigal,$^1$\thanks{E-mail: b.mcmonigal@physics.usyd.edu.au (BM)}
	N. F. Bate,$^1$
	G. F. Lewis,$^1$
	M. J. Irwin,$^2$
	G. Battaglia,$^{3,4}$
	\newauthor 
	R. A. Ibata,$^5$
	N. F. Martin,$^{5,6}$
	A. W. McConnachie,$^7$
	M. Guglielmo,$^1$
	A. R. Conn$^1$ \\
$^1$Sydney Institute for Astronomy, School of Physics, A28, The University of Sydney, Sydney, NSW 2006, Australia\\
$^2$Institute of Astronomy, Madingley Road, University of Cambridge, CB3 0HA, UK\\
$^3$INAF - Osservatorio Astronomico di Bologna, via Ranzani 1, 40127, Bologna, Italy \\
$^4$Istituto de Astrofisica de Canarias, calle via Lactea s/n, San Cristobal de La Laguna, 38200, Tenerife, Spain \\
$^5$Observatoire Astronomique, Universite de Strasbourg, CNRS, F-67000 Strasbourg, France\\
$^6$Max-Planck-Institut f\"{u}r Astronomie, K\"{o}nigstuhl 17, D-69117 Heidelberg, Germany \\
$^7$NRC Herzberg Institute of Astrophysics, 5071 West Saanich Road, Victoria, British Columbia V9E 2E7, Canada\\
}

\begin{document}

\date{Accepted --; Received --; in original form --}

\pagerange{\pageref{firstpage}--\pageref{lastpage}} \pubyear{2014}

\maketitle

\label{firstpage}

\begin{abstract}
We present deep optical photometry from the DECam imager on the 4m Blanco telescope of over 12 deg$^2$ around the Carina dwarf spheroidal, with complete coverage out to 1 degree and partial coverage extending out to 2.6 degrees. Using a Poisson-based matched filter analysis to identify stars from each of the three main stellar populations, old, intermediate, and young, we confirm the previously identified radial age gradient, distance, tidal radius, stellar radial profiles, relative stellar population sizes, ellipticity, and position angle. We find an angular offset between the three main elliptical populations of Carina, and find only tentative evidence for tidal debris, suggesting that past tidal interactions could not have significantly influenced the Carina dwarf. We detect stars in the vicinity of, but distinct to, the Carina dwarf, and measure their distance to be 46$\pm$2 kpc. We determine this population to be part of the halo of the Large Magellanic Cloud at an angular radius of over 20 degrees. 
Due to overlap in colour-magnitude space with Magellanic stars, previously detected tidal features in the old population of Carina are likely weaker than previously thought.

\end{abstract}


\section{Introduction}

The current $\Lambda$CDM cosmology explains the formation of large galaxies like the Milky Way (MW) via hierarchical structure formation, which involves the gradual build up of systems over cosmic time through the accretion of smaller systems across all scales \citep{White1978}. Evidence of this process can be seen throughout the Local Group, most notably in Andromeda with the Giant Stellar Stream \citep{Ibata2001}, a remnant of a major accretion event, and in the MW itself by the tidally disrupted Sagittarius dwarf galaxy \citep{Ibata1994}. It is clear that the epoch of galaxy formation is ongoing, and by looking at the population of dwarf galaxies around the MW, we can study the process occurring in resolved stellar structures.

The dwarf galaxy population of the MW has unsurprisingly been studied in more detail than any other (see \citealt{Tolstoy2009} for a recent review, and \citealt{McConnachie2012} for a recent consolidation of Local Group dwarf galaxy properties). Among these, the Carina dwarf galaxy is one of the best studied due to its interesting star formation history (SFH) and signatures of tidal disruption; some of the main properties of Carina from the literature are summarized in Table \ref{tab:main}.

\begin{table} 
  \centering
   \begin{minipage}{140mm}
  \caption{Main properties of Carina from the literature}
\label{tab:main}
\newcolumntype{C}[1]{>{\hsize=#1\hsize\centering\arraybackslash}X}
  \begin{tabularx}{80mm}{l C{1.5} C{0.5}}
  \hline
 Right Ascension (J2000.0) & 06h41m36.7s & (1) \\
 Declination (J2000.0) & $-50$d57m58s & (1) \\
 $l(^{\circ})$ & 260 & (1) \\
 $b(^{\circ})$ & $-22.2$ & (1) \\
 $\mu_{l}$(mas year$^{-1}$) & $-0.08$$\pm$0.09 & (2) \\
 $\mu_{b}$(mas year$^{-1}$) & 0.25$\pm$0.09 & (2) \\
 $D_{\odot}$(kpc) & $101\pm5$ & (1) \\
 $R_{peri}$(kpc) & 24$\pm$5 & (2) \\
 $V_{\odot}$(km s$^{-1}$) & 224$\pm$3 & (1) \\
 $V_{MW}$(km s$^{-1}$) & 20$\pm$24 & (3) \\
 Metallicity [Fe/H] & $-1.5$$\pm$0.2 & (4) \\
 Absolute Magnitude ($M_{V}$) & $-9.3$ & (4) \\
 Tidal Radius (arcmin) & 28.8$\pm$3.6 & (4) \\
 Ellipticity & 0.33 & (4) \\
 Position Angle$(^{\circ})$ & 65 & (4) \\
 \hline
\end{tabularx}
 \end{minipage}
 \noindent \textbf{References.} (1) \citet{Mateo1998}; (2) \citet{Penarrubia2009}; (3) \citet{Piatek2003}; (4) \citet{Irwin1995} \\

\end{table}

Carina was first argued to have have an age spread by \citet{Mould1983}, and is now known to have an unambiguously episodic SFH (e.g. \citealt{Smecker-Hane1994}; \citealt{Smecker-Hane1996}; \citealt{Mighell1997}; \citealt{Hurley-Keller1998}; \citealt{Monelli2003}; \citealt{Koch2006}; \citealt{Bono2010}; \citealt{Pasetto2010}). Short bursts of star formation separated by long pauses have created populations of ancient ($>$10 Gyrs), intermediate ($\sim$7 Gyrs), and young ($<$2 Gyrs) stars, as confirmed by the existence of multiple Main-Sequence-Turn-Offs (MSTOs) \citep{Hurley-Keller1998}. 
Such a punctuated SFH, while inferred for a number of isolated dwarfs such as I Zw 18 \citep{Annibali2013}, is difficult to explain if treated as an isolated system unless the system is near perigalacticon \citep{Nichols2014}; however Carina is thought to be near apogalacticon \citep{Piatek2003}, and if dynamically driven, we are led to infer a complex history of interactions.

The younger populations have a higher metallicity than the old population \citep{Koch2006}, 
so are unlikely to have formed by the accretion of new unenriched gas; they are more likely the result of re-accretion of gas blown out of Carina in the past \citep{Munoz2006}. \citet{Smecker-Hane1994} suggested that the gas could have been blown out by supernovae, but unless Carina has a very large dark matter (DM) halo, it would not be able to retain the gas. Normal star formation bursts are expected to only blow the gas out for $\sim$250Myr \citep{Salvadori2008}, and even if some mechanism did blow the gas out for $\sim$5 Gyr between each burst, again Carina would need to have a large DM halo in order for the gas to stay bound during the close pericentric passages with the MW, since the orbital period for Carina is estimated to be only 1.5-2 Gyr (e.g. \citealt{Lux2010}).

These scenarios are unlikely as such a large DM halo would also protect the stellar component of the dwarf from tidal interactions, yet there is abundant evidence of such interactions (e.g. \citealt{Munoz2006} and references therein), such as a profile break from a King to a power law (\citealt{Munoz2006}; \citealt{Majewski2005}). If this break was caused by tidal stripping, then at least at perigalacticon, Carina's DM halo must have been negligible \citep{Moore1996}, however the break may instead simply represent the boundary of efficient star formation \citep{Penarrubia2009}.

Other evidence of tidal interactions include the large spatial extent of `confirmed' member stars (e.g. \citealt{Munoz2006}; \citealt{Irwin1995}; \citealt{Kuhn1996}; \citealt{Monelli2004}); the orientation of the extended component along the major axis, with increasing ellipticity outwards \citep{Munoz2006}; and the flat dispersion profile which rises at larger radii \citep{Munoz2006}.

An analysis of recent CTIO/MOSAIC II deep field observations found evidence of tidal debris around Carina, with Carina members extending beyond the tidal radius \citep{Battaglia2012}. In this paper we seek to further examine the existence of such debris using deep optical photometry from the DECam imager on the 4m Blanco telescope. Using a Poisson-based matched filter analysis, we spatially map the main stellar populations separately to reveal only minor evidence of tidal debris. Stars from the Large Magellanic Cloud are also serendipitously detected and their distance measured. 

In Section \ref{sec:data} we discuss the DECam data used throughout this paper and present the fully calibrated spatial map. In Section \ref{sec:cmds} we discuss the key features of the Colour Magnitude Diagrams (CMDs) for each field. Section \ref{sec:match} introduces the matched filter method used in this paper, and Section \ref{sec:maps} discusses the maps of the different main stellar populations isolated by it. We discuss the implications of these measurements and conclude in Section \ref{sec:discussion}.

\section{Data}\label{sec:data}
Deep optical photometry of the Carina dSph was taken using the DECam imager mounted on the 4m Blanco telescope at the Cerro Tololo Inter-American Observatory (PI: G. F. Lewis). 
The large field of view of DECam allowed Carina to be imaged out to a radius of 1 degrees ($\sim$1.8kpc at the distance of Carina, approximately twice the nominal tidal radius; \citealt{Irwin1995}; \citealt{Mateo1998}) in a single pointing, and this was extended with 4 additional pointings of a 6 petal configuration centered at a radius of 1.5 degrees, such that two pointings are along the axis probed by \citet{Battaglia2012}, covering a total of over 12 square degrees. The final 2 petals will be observed at a future time, bringing the area covered to over 16 square degrees, and completing the coverage of Carina out to a radius of $\sim$2 degrees, which at the distance of Carina corresponds to $\sim$3.5kpc, roughly 4 times the nominal tidal radius.

Observations were taken in the $g$ and $r$ bands, reaching down to magnitudes of 24.9 and 24.6 respectively at 90 per cent completeness to make a clean star-galaxy separation in the regions of interest.
Each field was observed as a series of 500 second individual exposures (3 in $g$ and 5 in $r$), with a dither between exposures to reduce the impact of cosmic rays, field gaps and bad pixels.

The data were reduced within the Cambridge University Survey Unit (CASU), using a dedicated pipeline for working with wide-field camera data \citep{Irwin2001}; the pipeline was optimized for working with DECam data and performed standard reductions (debiasing, flat-fielding, astrometric and photometric calibration), as well as star-galaxy separation and the cataloging of sources. 

Foreground galactic extinction was corrected for using the \citet{Schlegel1998} infrared-based dust map, assuming the extinction coefficients of \citet{Schlafly2011}. 
Then stars were matched to entries in the Stetson Photometric Standard Fields catalog\footnote{http://www3.cadc-ccda.hia-iha.nrc-cnrc.gc.ca/community/STETSON/standards/} (\citealt{Stetson2000}; \citealt{Stetson2011}) based on astrometric data. 
The 4110 matches which had both $B$ and $V$ magnitudes, were transformed to AB $g$ and $r$ using the standard equations of Lupton\footnote{http://www.sdss.org/dr7/algorithms/sdssUBVRITransform.html}. These were then used to calibrate the zero-offset in each band.
During calibration, several exposures plagued by bad weather were identified as unusable, but their removal resulted in a uniform calibration across the survey region.

\begin{figure*}	
  \includegraphics[width=160mm]{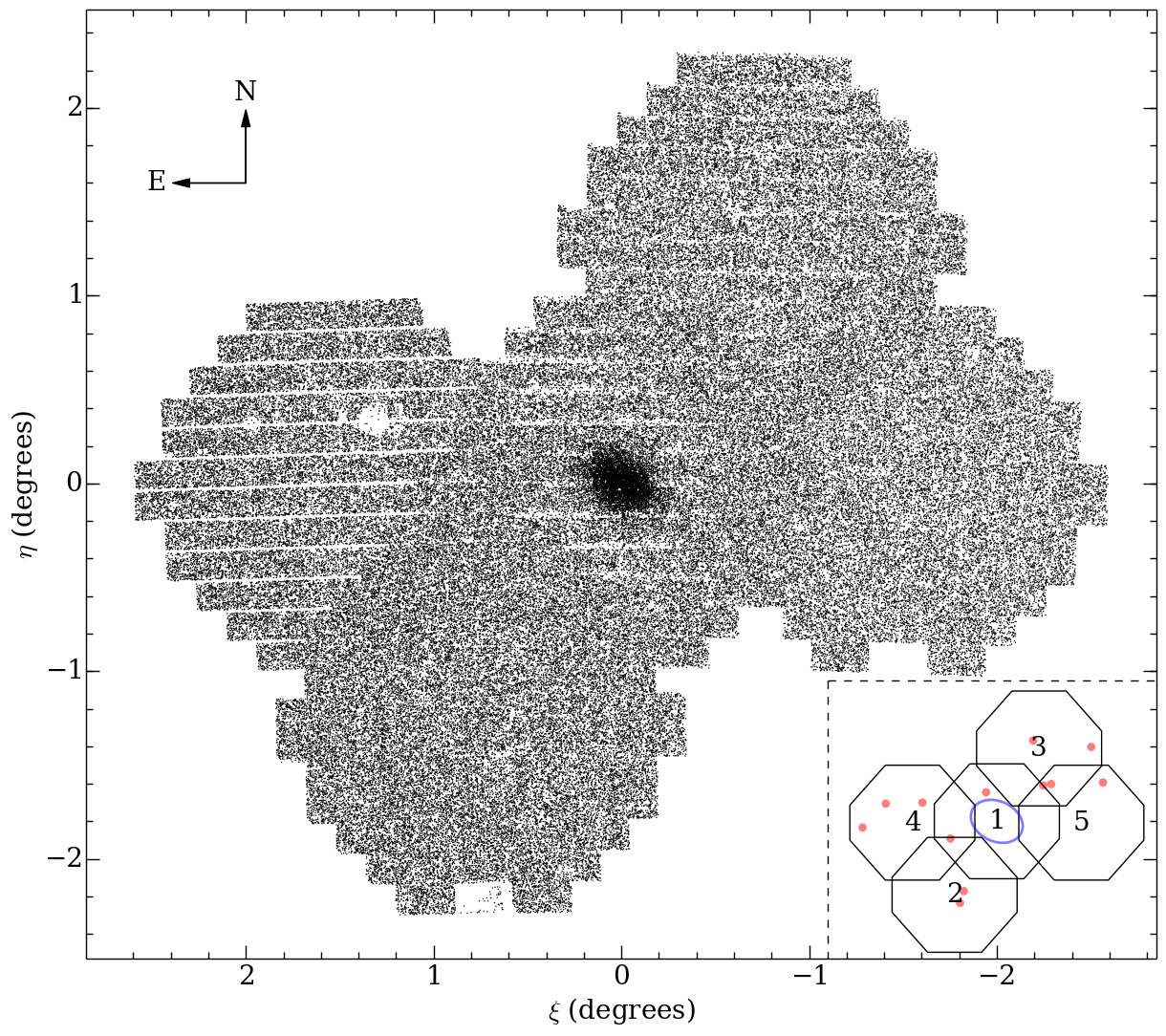}
  \caption{A tangent-plane projection centered on the Carina dwarf. All catalog entries identified as stellar and with $g_0$ and $r_0$ band values between 17 and 23.3 are plotted. The inset on the lower right shows the layout and general coverage of the fields, the tidal radius (blue), and the 12 brightest foreground stars (red).}
  \label{fig:distribution}
\end{figure*}

Figure \ref{fig:distribution} plots every star with $g_0$ and $r_0$ band values between 17 and 23.3, as a tangent-plane projection centered on Carina, along with a map of the fields. The Carina overdensity is clearly visible in the center. 
The CCD gaps are caused by a combination of reduced dither pattern coverage and the sensitivity of the morphological classifier to slight degradations in image quality around the outer parts of the stacked detector dither pattern due to seeing variations in the component images. The CCD gaps are most clearly visible in field 4, which was the most affected by weather, and are decreasingly noticeable in fields 1, 3, 5, and then 2. Field 1 had the largest variation in seeing for the component images which is the main reason for the gap visibility in this field. The CCD gaps are largely hidden in the regions where fields overlap leading to slightly better coverage in these regions. 

More significantly, z distortions in the (x-y) focal plane detector geometry resulted in some of the outer parts of the focal plane being at a different focus, particularly to the SW, than the rest of the focal plane. This gives a slight bias to the morphological classification in these outer regions. Similar isopleth maps constructed without any requirement on object morphology do not show this spatial pattern. However, without a constraint on object morphology the signal:to:noise of the map degrades significantly. We have experimented with accounting for this problem by stacking star count maps, for all bar the central field, to yield the generic underlying pattern and then applying this correction to all individual spatial isopleth maps. This mitigates against the main impact of the problem but still does not completely remove it. As an alternative we have also experimented with a more conservative use of the star/galaxy morphological classification but this also fails to completely correct the spatial bias, with the additional expense of worse signal:to:noise due to enhanced background galaxy contamination. 

Approximately 12 bright foreground stars are to blame for the remaining blemishes, with each conspicuous blemish in Figure \ref{fig:distribution} corresponding to a marked star on the Figure's inset.
Other than the aforementioned features, the covered region is visually homogeneous.

\section{CMDs}\label{sec:cmds}

\begin{figure*}	
  \includegraphics[width=160mm, natwidth=496, natheight=274]{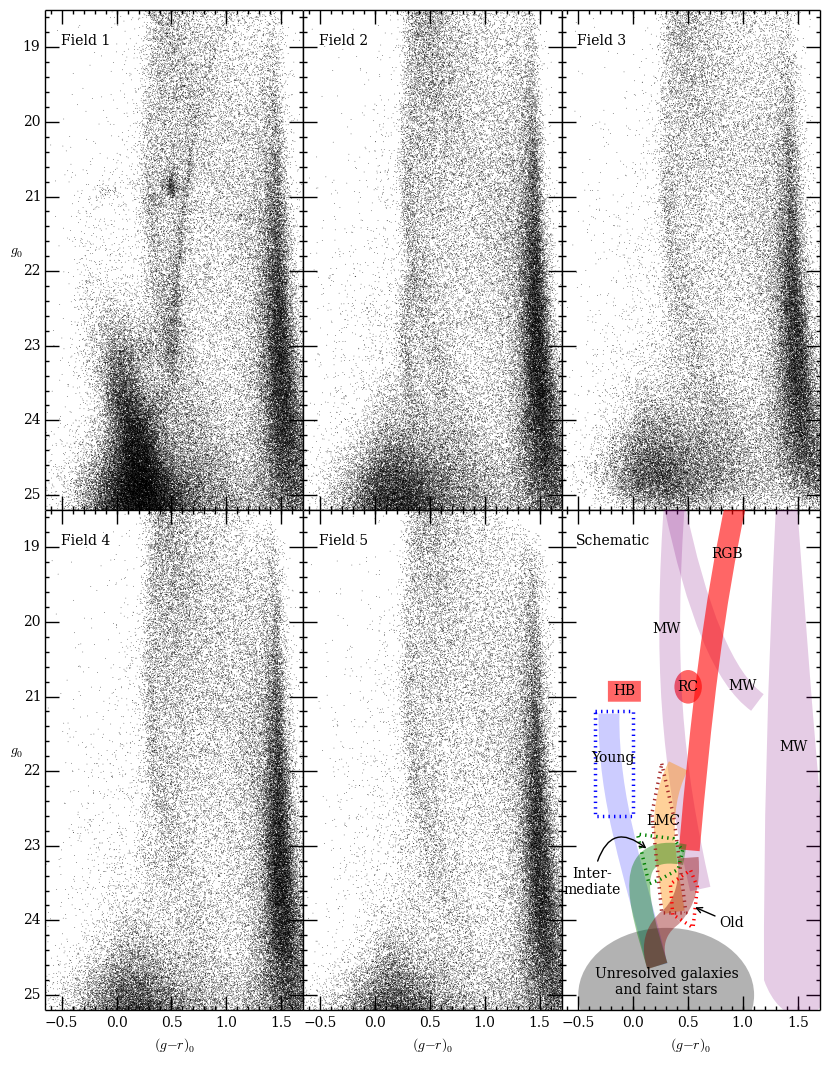}
  \caption{Colour Magnitude Diagrams (CMDs) constructed from all stars contained within each field. The lower-right panel plots a schematic representation of the main features of the CMDs and the location of the selection boxes used in the later analysis.}
  \label{fig:cmds}
\end{figure*}

Figure \ref{fig:cmds} shows the CMDs for each of the fields. The main sources of contamination in the catalog are clearly visible in each field as an overdensity at the bottom left due to mis-identified background galaxies, and an extended overdensity along the right-hand side due to MW foreground stars. Other contamination features visible in all fields are two foreground MW Main Sequences (MS), most prominently in field 2, running from (0.4, 17.5) to (1.3, 21.6) and from (0.4, 19.6) to (0.7, 24.0). Field 2 is also notable due to a bright Magellanic MS contaminant arcing from (0.4, 22.2) to (0.4, 23.9), passing through both the intermediate and old stellar population selection regions, with endpoints in the the dimmer MW MS contamination and the mis-identified galaxy overdensity; this is explored in Section \ref{sec:Magellanic}. Field 3 is notably not as deep as the other fields by 0.4 magnitudes; because of this, the old population selection box contains substantial contamination from mis-identified galaxies. In all other respects, fields 2, 3, 4 and 5 are equivalent.

\begin{figure*}	
  \includegraphics[width=160mm]{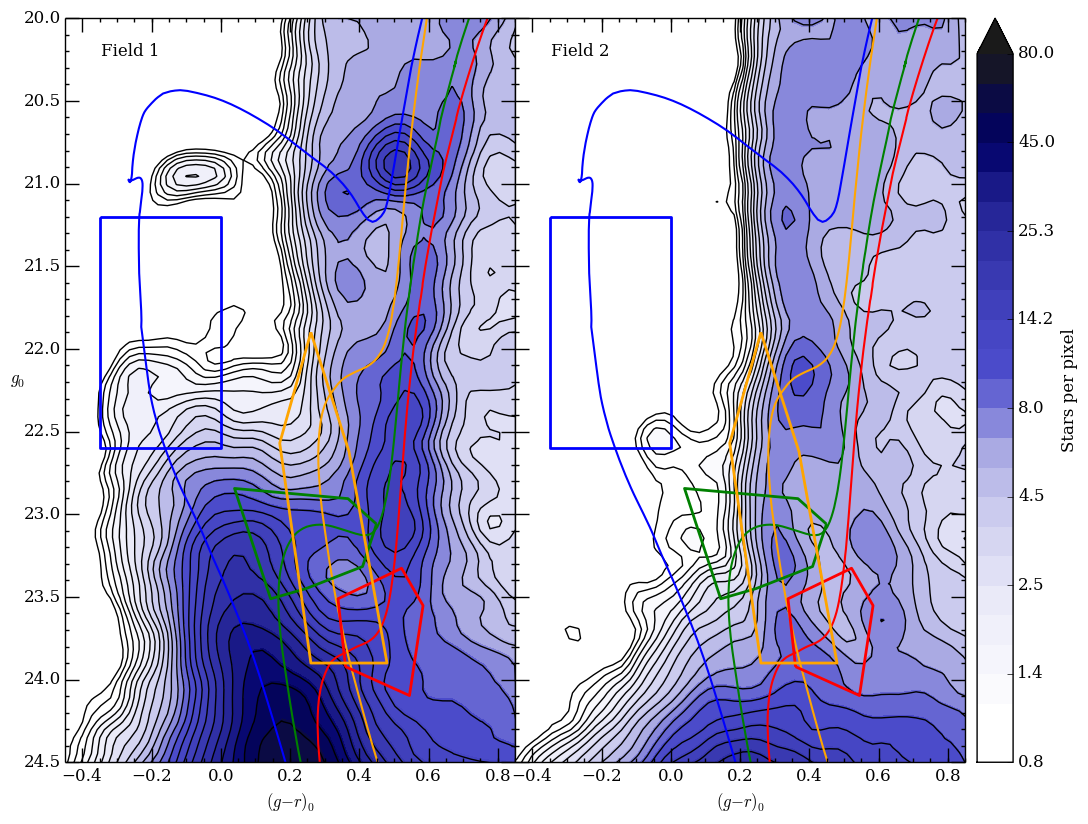}
  \caption{A magnified view of the CMDs for fields 1 and 2 constructed from all stars contained within each field with $g_0$ band values between 17 and 25. The maps were smoothed using a two-dimensional Gaussian with a 0.05 magnitude dispersion. The contours are spaced logarithmically, and the pixels cover 0.0013 square magnitudes. The best fitting isochrones for the main populations of Carina, young, intermediate, and old are coloured in blue, green, and red respectively. The selection boxes which best match these isochrones were chosen so as to lay on the locally densest parts of the diagram, without coming too close to each other, and have matching colour. The Magellanic feature, dominant in the right panel, is fitted with the orange isochrone. Its selection box, also in orange, could not avoid the Carina population selection boxes, and so was fitted independently. }
  \label{fig:cmd_12}
\end{figure*}

Field 1 is the only field with visible features of Carina; the left panel of Figure \ref{fig:cmd_12} shows a smoothed magnified view of the key features. 
The intermediate population, with its Main Sequence Turnoff (MSTO) marked by the green box, is clearly dominant, with its Red Giant Branch (RGB) extending towards the top of the CMD. The Red Clump is visible about half way up the RGB, with the Horizontal Branch (HB) extending to the left just below it. The blue box marks the faint young MS population, extending up towards the HB. The old population's MSTO is marked by the red box. 

SDSS isochrones were generated using the Dartmouth Stellar Evolution Database\footnote{http://stellar.dartmouth.edu/models/index.html} for the main populations of Carina with ages of 1, 7, and 13 Gyrs, with a metallicity of [Fe/H] = $-1.4$ giving the best fit to the data. These isochrones have a distance modulus of 20, corresponding to a distance of 100 kpc, as would be expected for Carina. The selection boxes for the Carina populations marked in Figure \ref{fig:cmd_12} were chosen to fit the isochrones, the contour map, and to avoid the other populations. An alternate set of selection boxes running down the Carina MS gave qualitatively similar results, but could not clearly distinguish the different populations.
The distance to Carina was checked using the Red Clump absolute magnitude calibration technique of \citet{Bilir2013}. Assuming a metallicity of [Fe/H] = $-1.3$ to $-1.5$, we get a distance of 98$\pm$11 kpc, consistent with expectations.

The right panel of Figure \ref{fig:cmd_12} is the same as the left panel but now for field 2. The Magellanic feature is now dominant, and is fit by an 11 Gyr isochrone with a metallicity of [Fe/H] = $-1.1$ \citep{Majewski2009} and a distance modulus of 18.3$\pm$0.1, corresponding to a distance of 46$\pm$2 kpc. The selection of this isochrone will be discussed further in Section \ref{sec:discussion}. The selection box for this feature was chosen to match the contour map, without overlapping with the MW MS at the top, or the contamination from galaxies at the base. The selection box for the Magellanic feature has a large overlap with the intermediate and old population selection boxes for Carina; this will be discussed in Section \ref{sec:maps}.

\section{Matched Filtering} \label{sec:match}

The Carina CMD shows three distinct phases of star formation: an old MSTO, an intermediate-age MSTO, and a young MS. 
We wish to determine the spatial extent of each of these three populations. In \citet{Battaglia2012} this was done by using a region away from the centre of Carina (see their Figure 3) to establish a uniform contamination, and subtracting this from Hess diagrams of the signal field. Isodensity contours were then plotted in units of the noise.

Here we apply an alternative technique, based on the matched filtering of \citet{Kepner1999} and especially \citet{Rockosi2002}. Matched filtering in this context is a technique where the observed data for an object (typically a globular cluster or dwarf galaxy CMD) is used to build a filter that describes the shape of the source population in colour-magnitude space. This filter, along with an appropriate contamination, is applied to regions where the signal is weak to pick out source populations from the noise in a statistically meaningful way (typically via a maximum likelihood determination).

The standard matched filtering technique as described in \citet{Rockosi2002} and \citet{Odenkirchen2003} assumes Gaussian statistics, however it is clear from Figure \ref{fig:cmds} that our CMDs are governed by Poissonian processes. We therefore construct a modified matched filtering technique as follows.

We describe the number of stars $\lambda_{c, m, \xi, \eta}$ in a given $(\xi, \eta)$ spatial pixel, with a given colour and magnitude $(c, m)$ using the following model:

\begin{equation}
\lambda_{c, m, \xi, \eta} = \alpha_{\xi, \eta}S_{c, m} + C_{c, m}.
\label{eqn:model}
\end{equation}

$S_{c, m}$ is the source filter, in the form of a probability density function in colour-magnitude space (see Section \ref{sub:filters} for details on constructing this probability density function). The contamination is described by $C_{c, m}$ (see Section \ref{sub:filters}), and $\alpha_{\xi, \eta}$ is the number of source stars in the spatial pixel $(\xi, \eta)$. It is this last, $\alpha_{\xi, \eta}$, that we wish to determine.

The algorithm we use to determine $\alpha_{\xi, \eta}$ is:

\begin{enumerate}
\item For a given spatial pixel $(\xi, \eta)$, construct an observed Hess diagram truncated to the region of interest in colour-magnitude space.
\item Set $\alpha_{\xi, \eta}=0$.
\item Calculate $\lambda_{c,m, \xi, \eta}$ according to Equation \ref{eqn:model}.
\item Determine the probability for the observed number of stars $n_{c, m, \xi, \eta}$ given $\alpha_{\xi, \eta}$ in each colour-magnitude pixel.
\item Calculate the probability for $\alpha_{\xi, \eta}$ over the entire colour-magnitude space.
\item Increment $\alpha_{\xi, \eta}$ by 1, and return to step (ii) until a maximum $\alpha_{\xi, \eta}$ is reached ($\sim250$).
\end{enumerate}

We calculate the probability for the observed number of stars $n_{c, m, \xi, \eta}$ given $\alpha_{\xi, \eta}$ using a Poisson distribution:
\begin{equation}
p(n_{c, m, \xi, \eta} | \alpha_{\xi, \eta}) = \exp(-\lambda_{c, m, \xi, \eta})\frac{\lambda_{c, m, \xi, \eta}^{n}}{n!}.
\end{equation}

The probability for $\alpha_{\xi, \eta}$ in a given spatial pixel is therefore the product of the individual probabilities at each pixel in colour magnitude space:
\begin{equation}
p(\alpha_{\xi, \eta}) = \prod{p(n_{c, m, \xi, \eta} | \alpha_{\xi, \eta})}.
\end{equation}
We take the maximum likelihood value of $\alpha_{\xi, \eta}$ to be the number of source stars in a given spatial pixel. In all cases, we find that the probability distributions for $\alpha_{\xi, \eta}$ are sharply peaked around the preferred value.

In principle, each of the three source populations -- young, intermediate, and old -- could be treated simultaneously using this process. In our case, our selection boxes (see Figure \ref{fig:cmd_12}, and Section \ref{sub:filters}) do not overlap in colour-magnitude space, and so we conduct the filtering process for each population separately.

\subsection{Matched Filtering Sample Output}

\begin{figure}  
  \includegraphics[width=55mm, angle=270]{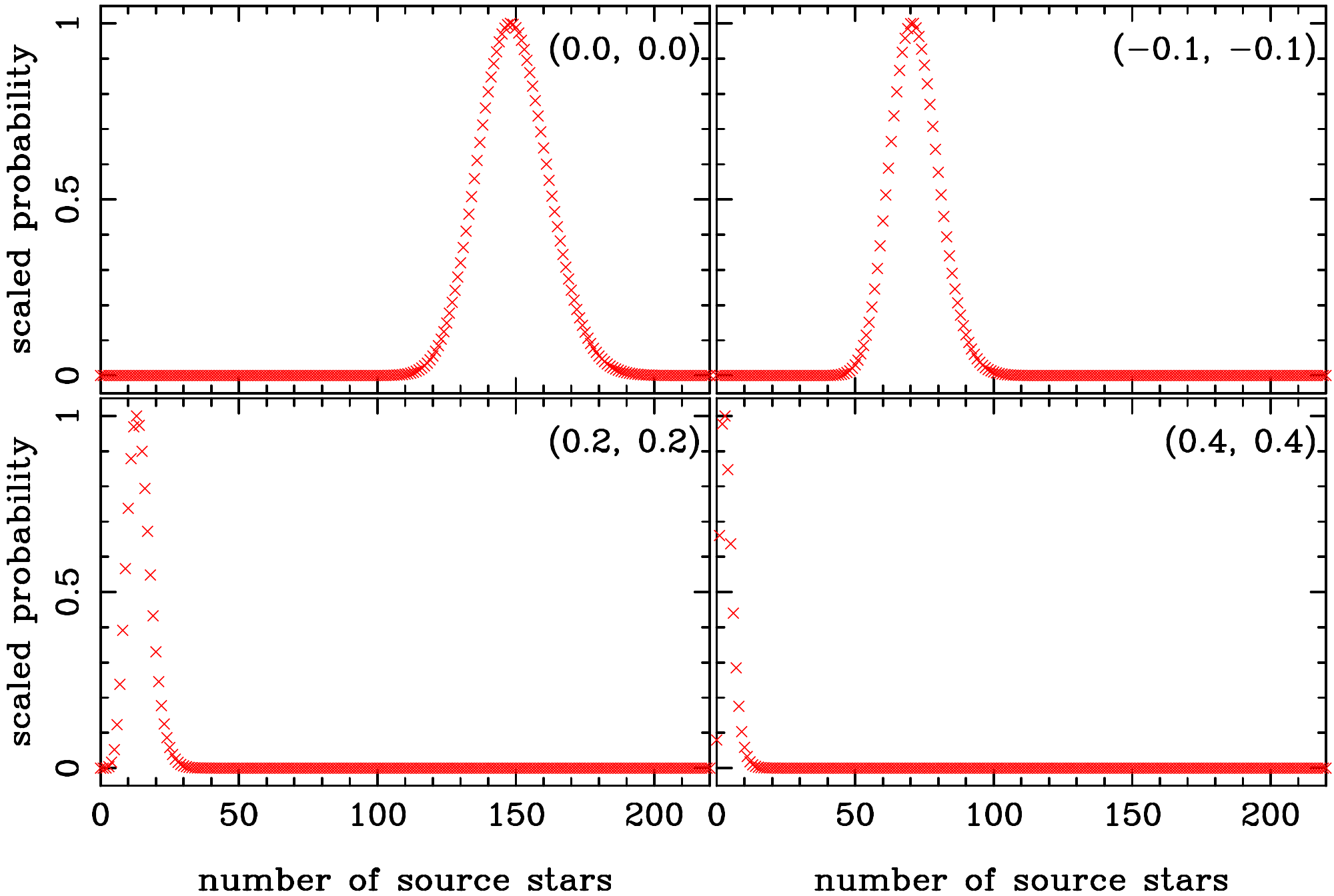}
    \caption{Sample output from the matched filtering algorithm for the full Carina population. Each panel shows the scaled probability distribution for the number of source stars $\alpha_{\xi, \eta}$ in a given $(\xi, \eta)$ pixel whose centre is displayed in the top right of each panel (in degrees). The top left panel shows the pixel in the centre of Carina. The next two panels show results running along the major axis to the South-West (top right) and North-East (bottom left). The bottom right panel is a pixel some distance from the centre of Carina towards the North-East, where the population of source stars has dropped to essentially zero. The most probable values of $\alpha_{\xi, \eta}$ in each pixel are used to generate the matched filtered maps displayed below.}
     \label{fig:probs_sample}
\end{figure}

Figure \ref{fig:probs_sample} shows some sample outputs for the matched filtering process. In all cases, the probability distribution for the number of source stars in a given spatial pixel $\alpha_{\xi, \eta}$ is simple. It is either piled-up at zero, or a single strong peak, making the interpretation of the probability distribution output straight-forward.

\subsection{Filters}
\label{sub:filters}
To generate source filters, a $(g_0-r_0, r_0)$ colour-magnitude diagram (CMD) was constructed from a circle with radius $10\arcmin$ centered on Carina, where the signal is strongest. This CMD was binned to create a Hess diagram with 0.025 magnitude pixels in both colour and magnitude, and smoothed using a two dimensional Gaussian with a dispersion of 2 pixels (0.05 magnitude). This is consistent with photometric errors.

Separate source filters for the old, intermediate and young populations were then generated by truncating the smoothed Hess diagram at each of the CMD boxes indicated in Figure \ref{fig:cmd_12}. Each filter was normalised independently so that the area under each was equal to one, thereby generating probability density functions for the old, intermediate and young populations in colour-magnitude space.

Contamination filters were generated by selecting regions well away from the centre of Carina. We chose four circles with $30\arcmin$ radius, one in each of our outside fields (see the red circles in Figure \ref{fig:large_maps}). A mean Hess diagram was constructed from the four fields, using 0.025 magnitude pixels in both $r_0$ and $g_0-r_0$, and then smoothed using a two-dimensional Gaussian with 0.05 magnitude dispersion. The resulting Hess diagram was normalized to unit area, and again truncated at each of the CMD boxes in Figure \ref{fig:cmd_12}.

\begin{figure*} 
  \includegraphics[width=150mm]{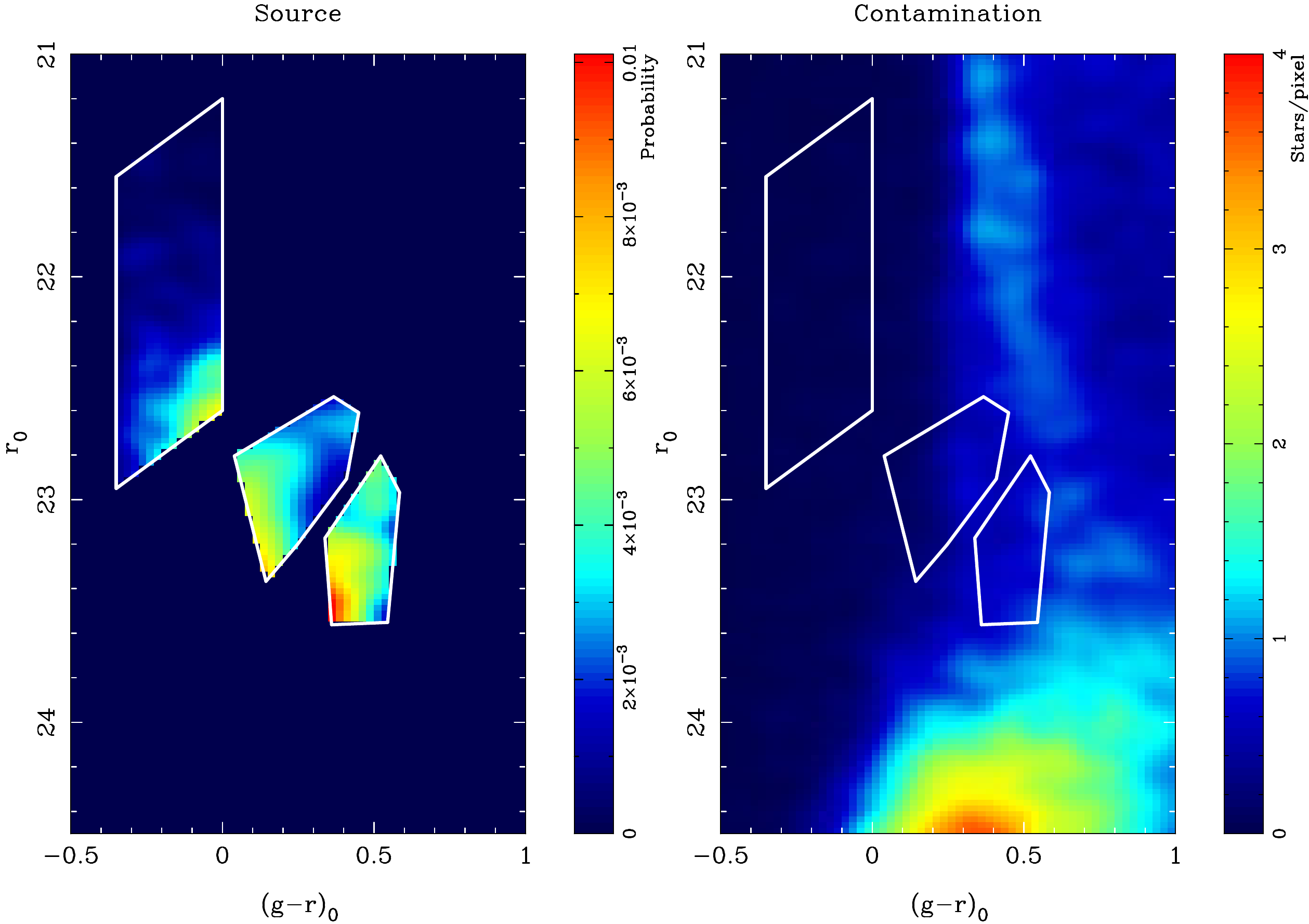}
    \caption{Filters used in the matched-filtering process. Left panel: source filters for the young, intermediate and old populations. These filters are probability density functions, and so each pixel represents the probability that a source star from that population has the given colour and magnitude. Right panel: contamination filters for the same three populations. These filters are normalised to unit area, and so represent an expected number of contaminant stars in each colour and magnitude pixel. Filters for all three populations are plotted together for convenience. Since the populations do not overlap, they are dealt with separately in the matched filtering process.}
     \label{fig:final_filters}
\end{figure*}

The final filters are provided in Figure \ref{fig:final_filters}. Source filters are probability density functions, thus each pixel represents the relative probability that a source star will have that pixel's colour and magnitude. The contamination filter, normalised to unit area, simply provides an expected number of contamination stars in each colour and magnitude pixel. To verify that our choice of contamination regions was not influencing our results, we repeated our analyses using contamination filters built from each outside field separately, and from every possible pairing of outside fields. There was no significant change in any of our matched filtered maps.

\section{Matched Filter Maps}
\label{sec:maps}
\begin{figure*}	
	\includegraphics[width=160mm]{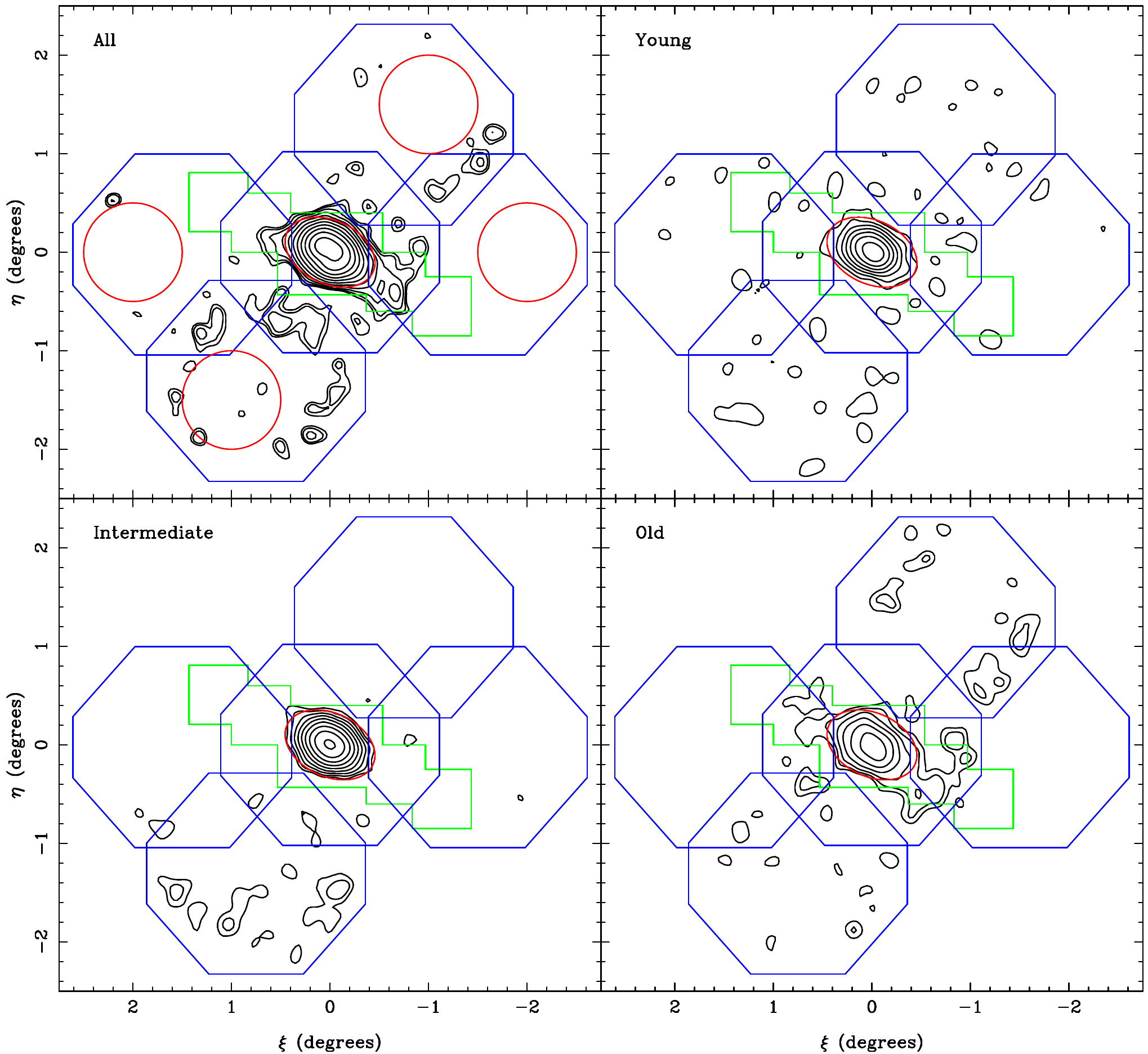}
  	\caption{Matched filtered stellar density maps for four stellar populations in Carina: the full population (top left panel), the young main sequence (top right panel), intermediate age main sequence turn off (bottom left panel), and old main sequence turn off (bottom right panel). The location of our five DECam fields are marked in blue, the survey region from \citet{Battaglia2012} in green, and the tidal ellipse in red (\citealt{Irwin1995}; \citealt{Mateo1998}). In each panel, the lowest contour is drawn at 2 times the RMS contamination level, determined in the four $30\arcmin$ circles marked in the top left panel. Subsequent contours are drawn non-linearly (see Section \ref{sec:maps}). Pixels are $2\arcmin\times2\arcmin$.}
  	 \label{fig:large_maps}
\end{figure*}

In Figure \ref{fig:large_maps} we provide spatial maps for four stellar populations in Carina, generated using the Poissonian matched filtering technique described above. These maps were generated on a $2\arcmin \times 2\arcmin$ pixel grid, and smoothed using a two-dimensional Gaussian with $4\arcmin$ dispersion. They show the full coverage of our DECam observations, with the separate DECam fields marked in blue. The region surveyed in \citet{Battaglia2012} is marked in green.

The lowest contour in each panel is drawn at 2 times the RMS contamination level, calculated using the contamination fields (marked with red circles in the top left panel of Figure \ref{fig:large_maps}). Below this level, there is signal throughout the observed fields. This level is 0.70 stars/pixel for the whole population (top left), 0.08 stars/pixel for the young population (top right), 0.26 stars/pixel for the intermediate population (bottom left), and 0.32 stars/pixel for the old population (bottom right). Subsequent contours are drawn non-linearly at $0.1 \times 1.75^{i-1}$ stars/pixel, where $i$ is the contour number. This avoids crowding in the central regions.

For matched filtering the entire Carina population (top left panels in the Figures), the following colour and magnitude cuts were used: $-1.0 < g_0 - r_0 < 1.1$ and $17.0 < g_0 < 23.3$. This cut most clearly separated the Carina CMD from sources of contamination.

\begin{figure*}	
	\includegraphics[width=160mm]{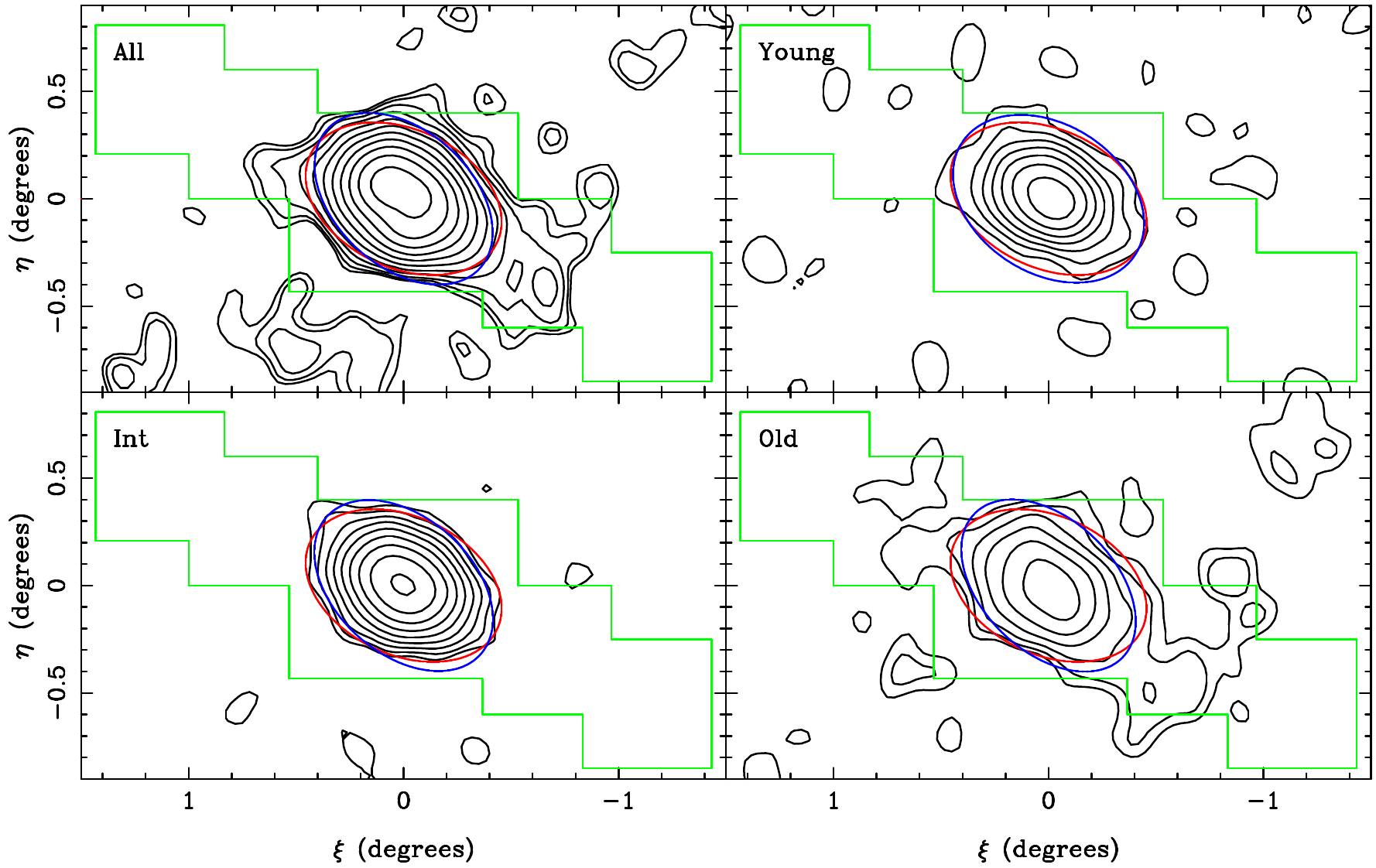}
  	\caption{Magnified view of the central regions of the Carina dwarf spheroidal. Four matched filtered maps are displayed: the full population (top left), young stellar population (top right), intermediate-age (bottom left), and old (bottom right). The \citet{Battaglia2012} survey area is marked in green, and the tidal ellipse in red (\citealt{Irwin1995}; \citealt{Mateo1998}). The blue ellipse in each panel is the corresponding ellipse fit to that population. Contours are drawn at twice the RMS contamination level (lowest contour), and incremented non-linearly, in each panel (see Section \ref{sec:maps}).}
  	 \label{fig:zoomed_maps}
\end{figure*}

In Figure \ref{fig:zoomed_maps}, we present a magnified view of the central regions of Carina. All other features of these maps are identical to those in Figure \ref{fig:large_maps}.

\subsection{Ellipticities and Position Angles}
Ellipticities and position angles are calculated by integrating over the spatially pixelated matched filtered maps to determine the quadrupole moments of the stellar densities; the results are summarized in Table \ref{tab:ellipse}. Uncertainties were calculated by by Poisson resampling the matched filtered maps 100,000 times. 

Going from the youngest population to the oldest population, the ellipticity increases and the major axis twists towards the horizontal axis, although the intermediate and old populations are consistent with having the same position angle. 
The full population closely matches the intermediate population, due to the numerical dominance of the intermediate population. While the literature value for the ellipticity approximately matches the value for the full population, the literature value of the position angle is only consistent with the young population, and is $\sim$15 degrees away from our values for the other populations.
We interpret the difference between the historical measurement and our measurement as a mild isophotal twist, noting that our moments based method is more sensitive to the inner regions.

\begin{table}
  \centering
   \begin{minipage}{140mm}
  \caption{Ellipticities and position angles}
\label{tab:ellipse}
  \begin{tabularx}{80mm}{lcc}
  \hline
   Population & Ellipticity & Position Angle (degrees) \\
 \hline
 Full & $0.34\pm0.01$ & $47.6\pm0.9$ \\
 Young & $0.28\pm0.02$ & $57.0\pm4.3$ \\
 Intermediate & $0.34\pm0.01$ & $48.1\pm1.7$ \\
 Old & $0.36\pm0.01$ & $45.9\pm1.3$ \\
 Literature & $0.33\pm0.05$ & $65.0\pm5.0$ \\

\end{tabularx}
 \end{minipage}
\end{table}

\subsection{Magellanic Feature}\label{sec:Magellanic}

\begin{figure*}	
	\includegraphics[width=160mm]{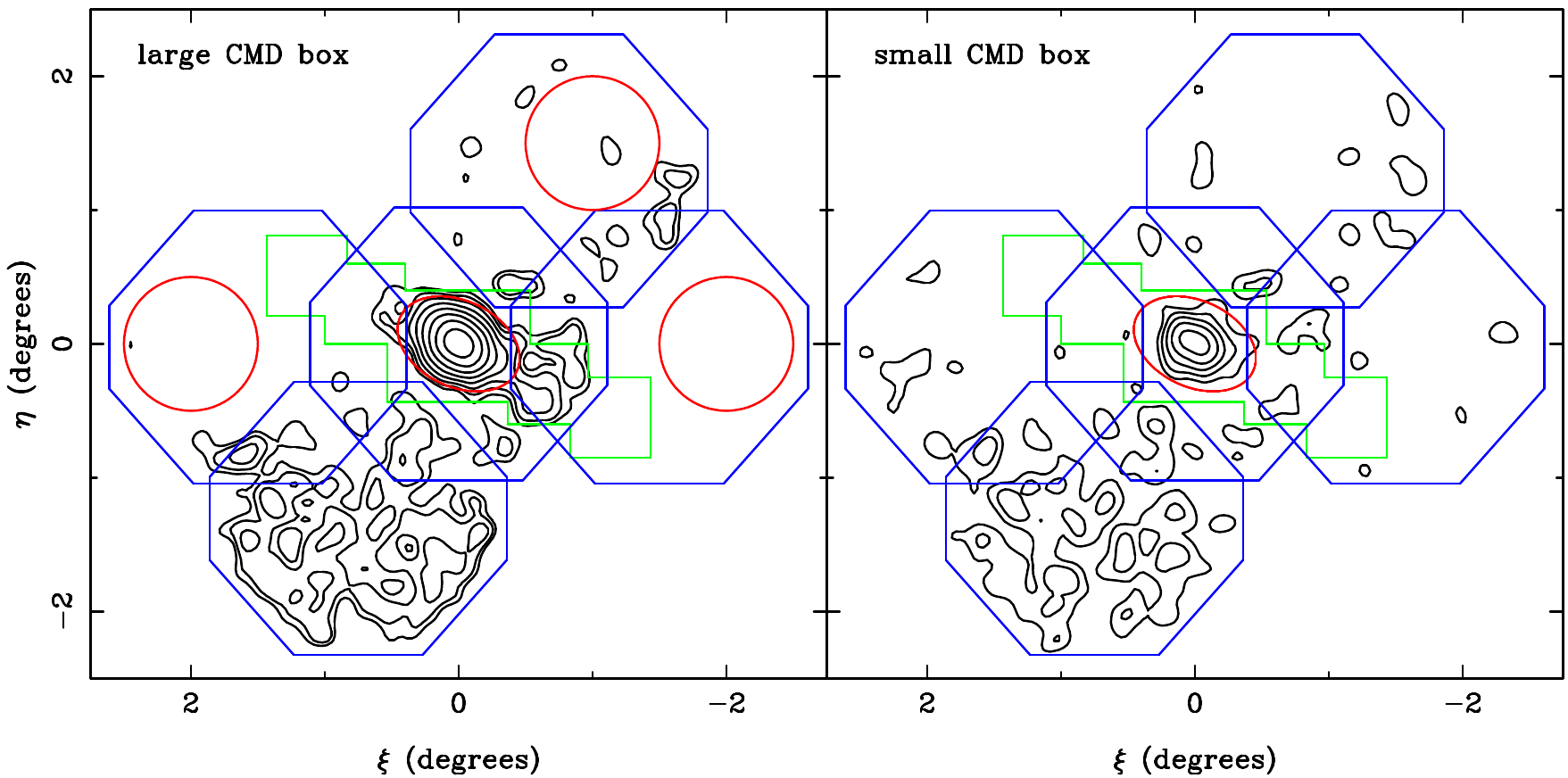}
  	\caption{Matched filtered stellar density maps for LMC stars. The left panel uses a large CMD box (see Figure \ref{fig:cmd_12}) that overlaps some known Carina source populations. The right panel uses a smaller CMD box that aims to exclude clearly identified Carina stars in the CMD. The tidal ellipse for Carina, our DECam fields and the \citet{Battaglia2012} survey region are marked in red, blue and green respectively. Pixels are $2\arcmin\times2\arcmin$, smoothed using a two-dimensional Gaussian with $4\arcmin$ dispersion. Contours are plotted linearly, from 2 times the RMS contamination level (calculated using the three red circles marked in the left panel) to 10 times the RMS contamination in steps of 1 times the RMS contamination.}
  	 \label{fig:lmc_maps}
\end{figure*}

In Figure \ref{fig:lmc_maps} we provide spatial maps for two selection boxes associated with the Magellanic feature in field 2. The larger selection box is the one shown in Figure \ref{fig:cmd_12}, selecting all stars clearly associated with this MS. The smaller selection box encompasses only the brighter stars from the larger box, so as to remove the bulk of the Carina members. The maps are largely the same, showing a fairly smooth distribution of stars throughout field 2. Carina shows up in both maps, as there is a large number of Carina stars in all parts of the CMD that the Magellanic feature covers.

In the left panel of Figure \ref{fig:lmc_maps}, there are a large number of stars directly to the South-West of Carina, matching the region in which the most tidal debris was found in \citet{Battaglia2012}. This region is mostly empty in the right panel, indicating that while it is still possible that these stars are actually Magellanic, they cannot be clearly separated from the intermediate and old Carina populations without metallicities or radial velocity data. The rough boundary of the Magellanic stars as taken from the left panel is approximately 20.7$^{\circ}$ from the Large Magellanic Cloud (LMC); this is consistent with the estimate of the boundary of the LMC halo in \citet{VanderMarel2014}.

\subsection{Radial Profile}\label{sec:radial}

\begin{figure*}	
	\includegraphics[width=160mm]{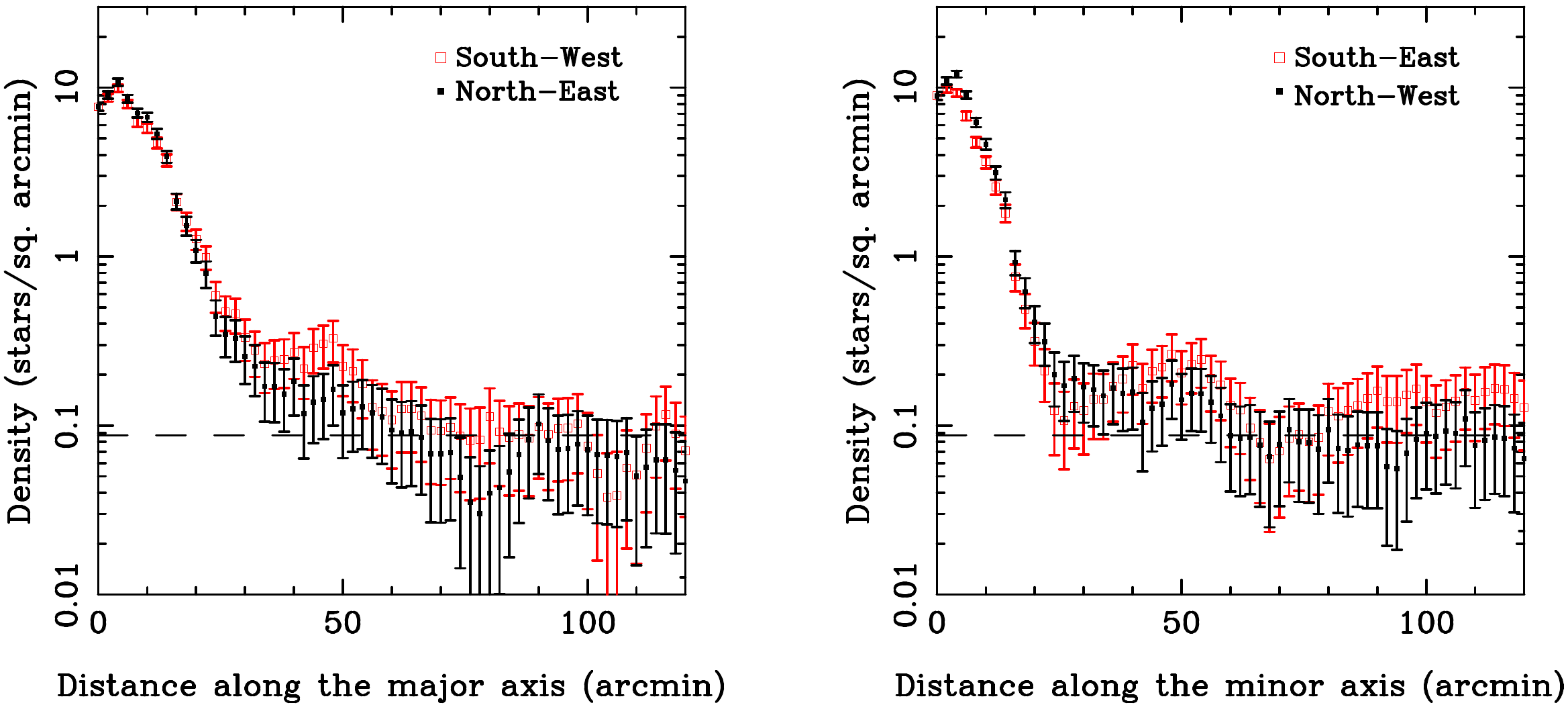}
  	\caption{Surface number count profile of stars in Carina, obtained from matched filtered maps of the entire Carina population (top left panel in Figures \ref{fig:large_maps} and \ref{fig:zoomed_maps}). Stars are contained in a $20\arcmin$ slice running along the \citet{Irwin1995} major axis (left panel), and perpendicular to the major axis (right panel). Open red squares denote stars to the South (West and East, left and right panels respectively), and filled black squares to the North (East and West, left and right panels). The dashed line marks the RMS stellar density across the matched filtered maps, used to determine contours in Figures \ref{fig:large_maps} and \ref{fig:zoomed_maps}. }
  	 \label{fig:radial}
\end{figure*}

Figure \ref{fig:radial} shows the stellar radial profiles obtained from matched filtered maps of the entire Carina population (top left panel in Figures \ref{fig:large_maps} and \ref{fig:zoomed_maps}) for the major and minor axes, using a $20\arcmin$ slice using the major axis calculated by \citet{Irwin1995} following \citet{Battaglia2012}. We see a small overdensity at $\sim$45 arc minutes South-West of Carina, and mild enhancements up to 60 arc minutes along all axes. 
We do not see an enhancement corresponding to the one found in \citet{Battaglia2012} around 70 arc minutes in any direction. It is worth noting that a small enhancement should be expected around 50 arc minutes in most directions, corresponding to the enhanced number density around the field edges as discussed in Section \ref{sec:data}; this may account for all but the largest overdensity already noted in the South-West. 
This overdensity seems to definitely be real, but it is not clear whether it is tidal debris from Carina, or misidentified Magellanic stars, as was suggested above for this location.

\begin{figure}	
	\includegraphics[width=80mm]{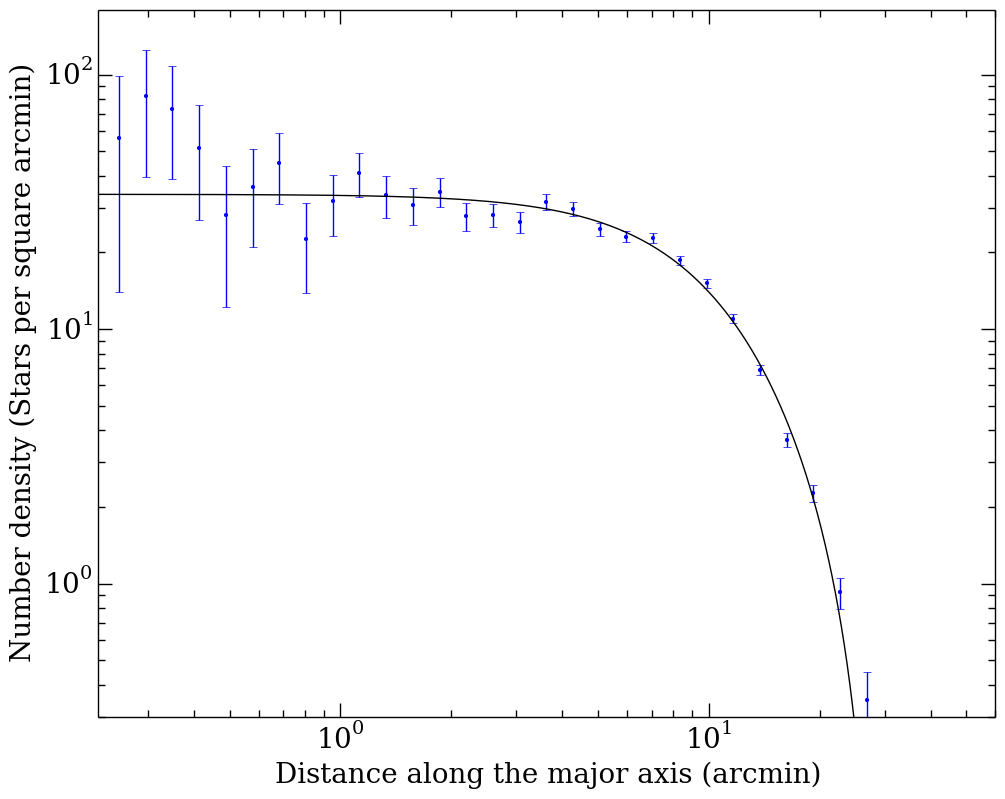}
  	\caption{Contamination subtracted surface number count profile of stars in Carina using stars with $17 < g_0 < 23.3$. 
Stars are binned into logarithmically spaced elliptical bins, using the ellipse from \citet{Irwin1995}. 
The solid line is the best fitting King profile \citep{King1962}. }
  	 \label{fig:tidal}
\end{figure}

A radial profile for all stars with $17 < g_0 < 23.3$, is shown in Figure \ref{fig:tidal}, using logarithmically spaced elliptical annuli based on the ellipse calculated by \citet{Irwin1995}. Following \citet{Irwin1995}, the best fitting King profile \citep{King1962} was found, giving a tidal radius of 29.1$\pm$0.9 arc minutes. Similar results in the range 30$\pm$2 arc minutes were found using the ellipse calculated from the full matched filter map, and the full matched filter map with each of these ellipses. All of these results are consistent with the tidal radius measured by \citet{Irwin1995}.

\subsection{Surface Brightness}

After contamination subtraction, the percentage of stars beyond the tidal radius out to a maximum of twice the tidal radius, as compared to the stars within the tidal radius is only 1.6 per cent. 
Carina has a magnitude of $M_{V} = -9.3$ \citep{Irwin1995}, so the total brightness of the tidal debris, including any misidentified Magellanic stars, is at most $M_{V} = -4.8$. Spreading this over the region between the tidal radius and twice the tidal radius, and assuming a distance modulus of 20.0, results in an average surface brightness of at most $\mu_{V} = 33.2$ magnitudes per square arc second for the tidal debris.

\section{Discussion and Conclusion} \label{sec:discussion}

We find the main features of Carina are in agreement with previous studies. Figure \ref{fig:zoomed_maps} shows the intermediate population is strongly dominant, and the general age gradient, with the young population being the most compact, and the old population being the most diffuse. Figure \ref{fig:radial} shows the radial profile of Carina along the major and minor axes, in general agreement with \citet{Battaglia2012}, but with a new overdensity to the South-West of Carina and without their main overdensity. The tidal radius was calculated to be 30$\pm$2 arc minutes. Ellipses were fit to each population, finding a gradient of ellipticities and major axis position angles with population age.

The Magellanic feature was fit by an 11 Gyr isochrone consistent with stars from the LMC. This gives a distance consistent with the LMC stars at this location in the sky \citep{Saha2010}. The LMC was shown to have a leading arm in the direction of Carina by \citet{Putman1998}, and subsequent LMC populations have been found around Carina by \citet{Majewski2000} and \citet{Munoz2006} who found LMC giants and LMC red clump stars respectively. 

Simulations of the formation of the stream find SMC stars \citep{Diaz2012} and LMC stars \citep{Besla2012} at the approximate location of Carina in the sky. Assuming that the contaminant stars in this study are LMC stars, there are several interpretations available. Either, they were stripped by an interaction with the MW, indicating that the LMC has a low mass $\sim10^{10}M_{\odot}$ or they could have been stripped by the SMC during the formation of the Magellanic Stream, which could allow for a larger LMC mass $\sim10^{11}M_{\odot}$ \citep{Besla2012}. It is also possible that we are seeing either a common envelope of the SMC-LMC pair, or simply that we have detected the edge of a very large diffuse LMC halo, which would indicate a high LMC mass.
In any case, potential interaction between Carina and the Magellanic clouds is a very complicated issue, and we therefore refrain from further speculation.

Figure \ref{fig:large_maps} shows substructure at twice the RMS level to be evenly distributed across the entire map in the young population, which is consistent with noise. The intermediate population is stronger, so the map is cleaner, with the stars being largely centrally located. The noise throughout field 2 is the result of the unavoidable overlap with the Magellanic feature. The old population is both weak, and deep in the CMD, close to the noise level, so has the lowest quality map. There is some signal in field 2, again due to the overlap with the Magellanic feature, but there is also signal around the North-East and South-West of field 3; this is likely due to noise, and is also present in the Magellanic feature matched filtered maps. The signal around Carina, strongest to the South-West is likely a combination of a small amount of tidal debris, and some Magellanic stars. Unfortunately it is not possible to differentiate between them with this data.

Although Carina has a high heliocentric radial velocity, in a galactocentric frame the radial velocity is close to zero. In the likely event that Carina is not on a circular orbit about the Milky Way this suggests that it is either currently at apocentre or pericentre. \citet{Piatek2003} used their measured HST proper motion to suggest that Carina is currently at apocentre and approaches to within $\approx$20\,kpc at pericentre. However, the HST proper motion errors for Carina, anchored by a single Quasi Stellar Object as reference, are large enough that we can not completely rule out the possibility that Carina could equally well be at pericentre. The presence or absence of clear signs of tidal disturbance, or tidal streams, is then quite pertinent. 

Furthermore, Carina has been impressively successful at retaining its gas over the best part of a Hubble time. If the orbital period is only 1-2 Gyr, as suggested by \citet{Piatek2003}, and if Carina plunges to within 20\,kpc of the Milky Way centre, then tidal effects coupled with ram pressure stripping of the gas have been singularly unsuccessful. On the other hand, if Carina is actually currently at pericentre, the complex star formation history coupled with the lack of clear signs of tidal disruption in the DECam data make more sense.

The apparent absence of Carina's tidal tails presents a challenge for the MOND simulations of \citet{Angus2014}, which find tidal tails as a natural result even for near circular galactic orbits.
We have found a much lower level of tidal debris than \citet{Battaglia2012}, which is consistent with our identification of the Magellanic stellar population conflated in colour-magnitude space. 
This data has gone a long way in aiding our understanding the Carina system, but further spectroscopic observations to obtain detailed metallicties and kinematics are required to complete this picture.

\section*{Acknowledgments}
NFB and GFL thank the Australian Research Council (ARC) for support through Discovery Project (DP110100678). GFL also gratefully acknowledges financial support through his ARC Future Fellowship (FT100100268). BM acknowledges the support of an Australian Postgraduate Award.
This project used data obtained with the Dark Energy Camera (DECam), which was constructed by the Dark Energy Survey (DES) collaborating institutions: Argonne National Lab, University of California Santa Cruz, University of Cambridge, Centro de Investigaciones Energeticas, Medioambientales y Tecnologicas-Madrid, University of Chicago, University College London, DES-Brazil consortium, University of Edinburgh, ETH-Zurich, University of Illinois at Urbana-Champaign, Institut de Ciencies de l'Espai, Institut de Fisica d'Altes Energies, Lawrence Berkeley National Lab, Ludwig-Maximilians Universitat, University of Michigan, National Optical Astronomy Observatory, University of Nottingham, Ohio State University, University of Pennsylvania, University of Portsmouth, SLAC National Lab, Stanford University, University of Sussex, and Texas A$\&$M University. Funding for DES, including DECam, has been provided by the U.S. Department of Energy, National Science Foundation, Ministry of Education and Science (Spain), Science and Technology Facilities Council (UK), Higher Education Funding Council (England), National Center for Supercomputing Applications, Kavli Institute for Cosmological Physics, Financiadora de Estudos e Projetos, Fundação Carlos Chagas Filho de Amparo a Pesquisa, Conselho Nacional de Desenvolvimento Científico e Tecnológico and the Ministério da Ciência e Tecnologia (Brazil), the German Research Foundation-sponsored cluster of excellence ``Origin and Structure of the Universe'' and the DES collaborating institutions.

\bibliography{Carina}

\bsp

\label{lastpage}

\end{document}